\documentclass[11pt,twoside]{article}
\usepackage{asp2004}
\begin{document}
\title{The Current Status of Primary Distance Indicators}
\author{Michael Feast}
\affil{Astronomy Dept,University of Cape Town, Rondebosch, 7701,South Africa\\
mwf@artemisia.ast.uct.ac.za}
\begin{abstract}
A review is given of the current status of the primary distance indicators.
The relevance of these indicators for determining the local expansion
rate and the age of globular clusters is briefly outlined.
\end{abstract}
\section{Introduction}
 In the cosmological context, primary distance indicators are of
importance as the calibrators of those more general indicators
that determine the local expansion rate and deviations from it.
They are also of importance in estimating the ages of globular clusters
and hence a lower limit to the age of the Universe.
  This talk has the quite limited objective of giving an overview of
the current calibration  of
primary distance indicators, 
and particularly trying to draw attention to present uncertainties.
Some implications of this work will then be briefly discussed.
\section{Cepheids}
The best known primary distance indicators are of course the Cepheids.
The slope of the Period-Luminosity Relation in the V band,
PL(V), has generally been taken from the LMC Cepheids. However the zero-point, 
and hence the scale, can be set by observations in our own Galaxy. 
Table 1  lists four ways of doing this Galactic calibration and gives
the corresponding zero-points, $\gamma$, of the PL(V) relation;
\begin{equation}
M_{V} = -2.81 \log P + \gamma                         
\end{equation}

\begin{table}[!ht]
\caption{~~~~~~~~~~~~~~~~~The Cepheid Zero-point}
\smallskip
\begin{center}
{\small
\begin{tabular}{lll}
\tableline
\noalign{\smallskip}
~~Method & ~~~~~~~$\gamma$ & ~~~~$\Delta$ \\
\noalign{\smallskip}
\tableline
\noalign{\smallskip}
Trig. Parallax & $-1.36 \pm 0.08$ & $-0.01$ \\
Stat. Parallax & $-1.46 \pm 0.13$ & $-0.11$ \\
Puls. Parallax & $-1.31 \pm (0.04)$ & $+0.04$ \\
Clusters       & $-1.45 \pm (0.05)$ & $-0.10$ \\
 & & \\
Unweighted (4) & $-1.40$ & \\
 & & \\
NGC4258 & $-1.17 \pm 0.13$ &  $+0.18$ \\
 & &\\
 & & \\
Unweighted (5) & $-1.35 \pm (0.05)$  & \\
\noalign{\smallskip}
\tableline
\end{tabular}
}
\end{center}
\end{table}

The straight forward and fundamental way is to use trigonometrical parallaxes
of nearby Cepheids. The value in the Table 1 combines Hipparcos parallax
results (Feast \& Catchpole 1997) 
with a parallax of delta Cephei itself obtained by Benedict
et al. using the HST (see Benedict et al. 2002b, Feast 2002). 
A group led by Fritz Benedict has an HST project in
progress to determine the parallaxes of 10 other local Cepheids. 
It is hoped that the relative
parallax errors of the 10 stars, which is 35 percent in Hipparcos
will be reduced to about 8 percent. The stars cover a range in period
and so should give us some hold on the PL(V) slope in our Galaxy.

Before the Hipparcos improvements of Cepheid parallaxes, the most favoured
way of fixing the Cepheid scale was from Cepheids in open clusters,
using distances from main sequence fitting. This method has a number of
problems. For instance the steepness of the main sequence makes the method
sensitive to the adopted reddening and also to photometric errors. 
The standard method has been to use the Pleiades as a template with some 
adopted Pleiades modulus. Although the individual Hipparcos
parallaxes of Pleiades stars are relatively poor due to its
considerable distance, 
van Leeuwen (1999, 2000) and Robichon et al. (2000) obtained
a mean parallax of good internal consistency by combining
all the Hipparcos results for stars in the cluster.
The distance derived was considerably smaller than expectations. 
In view of doubts regarding these results
it has seemed prudent to use as a standard the distance
of the Hyades which is much closer and for which good Hipparcos
parallaxes of individual stars were obtained
(Perryman et al. 1998). It is a simple matter to derive
from main sequence fits the relative Hyades/Pleiades distances. It does
however require a correction for the Hyades metallicity, though this is 
believed to be well known and this leads to the cluster result in Table 1
(Feast 2000, 2003).

It is worth  discussing briefly the ``Pleiades problem"  and to see
what its implications are for Hipparcos parallaxes in general. 
Makarov (2002, 2003) has investigated this problem in some detail. 
These papers
suggest that the Hipparcos reference frame suffers from slight random
errors which, in a small area of the sky, may be correlated to some extent.
The uncertainty of the mean parallax of N stars in such an
area may then not decrease as root N. By going
back to the original Hipparcos data Makarov shows how this problem can
be at least partially overcome. Makarov's result leads to 
the Pleiades modulus given in Table 2. 

\begin{table}[!ht]
\caption{~~~~~~~~~~~~~~~~~~~~The Pleiades Modulus}
\smallskip
\begin{center}
{\small
\begin{tabular}{ccc}
\tableline
\noalign{\smallskip}
Method & Modulus & $\Delta$ \\
\noalign{\smallskip}
\tableline
\noalign{\smallskip}
Original Hipparcos & $5.37 \pm 0.07$ & \\
------------------ &   --------- & \\
 & & \\
Revised Hipparcos & $5.57 \pm 0.06$ & $-0.05$\\
Atlas & $5.67 \pm 0.04$ & $+0.05$ \\
HD 23642 & $5.60 \pm 0.04$ & $-0.02$ \\
Via Hyades & 5.66 & $+0.04$ \\
 & & \\
Unweighted (4) & 5.62 & \\
\noalign{\smallskip}
\tableline
\end{tabular}
}
\end{center}
\end{table}

Whilst a problem of this kind may be important for some clusters,
Makarov's work suggests that for
objects scattered over the sky this type of error 
will be randomly distributed and the 
individual parallaxes of such objects can be
combined satisfactorily
in the normal way. Whether further work will establish that the
overall standard errors of Hipparcos stars need to be slightly increase
is not yet clear. Even if they did, the method of reduced parallaxes
which is used for instance in reducing the Cepheid data (e.g.
Feast \& Catchpole 1997),
only relies on the relative values of the 
absolute errors of the stars involved and since these vary little 
from star to 
star the effect seems likely to be small, though of course it would
be desirable to have numerical confirmation of this.
 
Besides the revised Hipparcos modulus of the Pleiades, Table 2 contains
two other recent estimates. 
Pan et al.(2004) obtain a distance from an interferometric orbit of the 
binary member,
Atlas, together with an adopted mass-luminosity relation. 
Another estimate is  found from the eclipsing binary member, HD23642
(Munari et al. 2004), adopting a colour-surface brightness relation.
Table 2 shows that all the current values, including the result obtained
via the Hyades, which has been adopted in the present discussion, agree well
($\Delta$ is the residual from the final mean).

There are two other methods of fixing the galactic Cepheid scale shown in
Table 1; statistical parallaxes and pulsation parallaxes.
The main uncertainty in the case of statistical parallaxes is the need to 
adopt a model of galactic motions. Fortunately in the case of the Cepheids
distributed over a large volume, their motions are dominated by
differential galactic rotation. This is independently demonstrated
 by both proper motions and radial velocities. The radial velocity
data can then be easily scaled to fit the proper motion results and to
yield the scale. The zero-point quoted uses Hipparcos proper motions and
is from Feast \& Whitelock (1997).
There has been extensive further work on stellar proper motions generally,
for instance the current US Naval Observatory catalogue
(Zacharias et al. 2004), and it would be
worthwhile investigating whether they 
contain data which could be used to improve the
zero-point.

There is much work at present on various forms of the pulsation
parallax method. These include the current ability to determine
the angular diameters of Cepheids, and their variation with phase, using
interferometry 
(see for instance, Fouque, Storm \& Gieren 2003 and references there, also
Kervella et al. 2003, 2004a,b). 
Pulsation parallaxes can give results of high
internal consistency. However it is quite difficult to estimate
possible systematic uncertainties in combining radial velocities with
surface brightness or angular diameter measures,  due to limb
darkening, atmospheric complexities etc.
(see for instance, Marengo et al. 2003a,b). The value adopted here is from
Laney (1998) as representative

I have included in Table 1 the zero-point derived using Cepheids in NGC4258
whose distance is derived from the motions of $\rm H_{2}O$ masers round
a central black hole --together with a model
(Herrnstein et al. 1999, Newman et al. 2001). The metallicity of this
galaxy is close to solar, so only a small metallicity correction was
required. The deviation from the mean  ($\Delta$) is 
not significant.

A good deal of the current effort on Cepheids is being put into 
investigating in detail whether the PL relation is exactly linear and
how its slope and zero-point are affected by metallicity
(e.g. Gieren et al. 1998, 
Tammann et al. 2003, Sandage et al. 2004, Kennicutt et al. 1998,
Groenewegen et al. 2004, Storm et al. 2004, Sakai et al. 2004).
The slope has in the past been taken from the metal-poor LMC Cepheids. 
Both pulsation parallaxes and Cepheids in clusters suggest that for
galactic Cepheids the slope in PL(V) is slightly different. This is important
because the weighted mean abundance of the HST key project galaxies
is  near solar. Thus for instance using the galactic slope of
Gieren et al (1998) from pulsation parallaxes, rather than the LMC slope,
together with the trigonometrical  parallaxes would lead to an increase in the
HST key project distance scale of about 7 percent and a corresponding
decrease in $\rm H_{o}$, due to the difference in mean period of the
calibrating and programme Cepheids. This gives some idea of the
current uncertainties.

\section{The RR Lyrae Variables}
   Table 3 lists estimates of the zero-point ($\rho$) of the RR Lyrae, absolute
magnitude - metallicity relation;
\begin{equation}
M_{V} = 0.2([Fe/H] +1.5) +  \rho
\end{equation}
The adopted slope is a compromise between various suggested values (see e.g.
Gratton et al. 2003).

\begin{table}[!ht]
\caption{~~~~~~~~~~~~~~~~The RR Lyrae Zero-point}
\smallskip
\begin{center}
{\small
\begin{tabular}{ccc}
\tableline
\noalign{\smallskip}
Method & $\rho$ & $\Delta$ \\
\noalign{\smallskip}
\tableline
\noalign{\smallskip}
Trig. Par. & $0.57 \pm 0.11$ & $~~ 0.00$ \\
Hor. Branch & $0.56 \pm 0.15$ & $-0.01$ \\
Globulars & $0.56 \pm 0.07$ & $-0.01$ \\
$\delta$ Sct &  $0.44 \pm 0.10$ & $-0.13$ \\
Stat. Par. & $0.74 \pm 0.13$ & $+0.17$ \\
 & & \\
Unweighted (5) & 0.57 & \\
\noalign{\smallskip}
\tableline
\end{tabular}
}
\end{center}
\end{table}

The parallax result (Feast 2002) is from the HST parallax of RR Lyrae
itself determined by Benedict et al. (2002a). 
In principle the HST could be used to obtain parallaxes of other RR Lyraes 
which would substantially improve this estimate.
The statistical parallax result 
(Gould \& Popowski 1998) is based on a simple model of
galactic halo kinematics which is questionable in view of the possible
effects of streams. The other entries in Table 3 are from horizontal
branch stars with Hipparcos parallaxes
(Gratton 1998), from globular clusters with
distances from Hipparcos subdwarfs
(Gratton et al. 2003), and via the Hipparcos parallaxes of
$\delta$ Sct stars
(McNamara 1997). An unweighted mean of all the estimates is given in Table 3.

\section{The Mira Variables}

The infrared (K) PL relation for Miras was established in the LMC. It
has a small scatter (Feast et al. 1989) and avoids the reddening problem which is
significant in the optical region. 

An initial calibration of this relation using Hipparcos parallaxes
(Whitelock \& Feast 2000) has recently been revised by Whitelock
(to be published)
taking into account the chromatic corrections to the parallaxes
that have been suggested (Knapp et al. 20003, Platais et al. 2003
Pourbaix et al. 2003). Both the original and revised result are shown
in the Table 4 (note that the small bias corrections discussed in
Feast (2002) have been applied).

\begin{table}[!ht]
\caption{~~~~~~~~~~~~ The Mira PL(K) Zero-point}
\smallskip
\begin{center}
{\small
\begin{tabular}{ccc}
\tableline
\noalign{\smallskip}
Method & $\kappa$ & $\Delta$ \\
\noalign{\smallskip}
\tableline
\noalign{\smallskip}
Original Hipparcos Par. & $0.86 \pm 0.14$ & \\
----------------------- &  ----------- & \\
 & & \\
Revised Hipparcos Par. & $1.06 \pm 0.13$ & $+0.06$ \\
OH VLBI & $1.01 \pm 0.13$ & $+0.01$ \\
Globular Clusters & $0.93 \pm 0.14$ & $-0.07$ \\
 & & \\
Unweighted (3) & $1.00 \pm 0.08$ &   \\
\noalign{\smallskip}
\tableline
\end{tabular}
}
\end{center}
\end{table}

Also in Table 4 are the zero-point derived using the distances of a few
Miras obtained from VLBI of their OH masers  
(Vlemmings et al. 2003) and via metal-rich
globular clusters which contain Miras and have Hipparcos sub-dwarf
distances (Feast et al. 2002). The agreement between the various methods 
($\Delta$ in Table 4)
is better than one could reasonably expect.

The power of this method of distance determination was recently
demonstrated by Rejkuba (2004) who used the K magnitudes and periods
of about 1000 Miras in NGC5128 (Cen A) to obtain a distance for this
galaxy. Her results indicate that the PL(K) slope is closely the
same in the LMC and in NGC5128 and that based on the LMC distance
modulus adopted below (Table 5), the Mira modulus ($28.0 \pm 0.2$)
agrees with that found from the RGB tip ($27.9 \pm 0.2$)
\footnote{A discussion of the RGB tip is not given here since although
it seems a good distance indicator its calibration depends either on
other (primary) distance indicators or theory.}.

\section{Intercomparison of Primary Distance Indicators}
   It has often been suggested that the distance of the LMC is of
fundamental importance in the distance scale problem.  However the
basic zero-points are best established in Our Galaxy. Nevertheless
the LMC is of importance for comparing distance indicators with one another.

\begin{table}[!ht]
\caption{~~~~~~~~~~~~~~~~~~~~~~~~LMC Modulus}
\smallskip
\begin{center}
{\small
\begin{tabular}{ll}
\tableline
\noalign{\smallskip}
Method & Modulus \\
\noalign{\smallskip}
\tableline
\noalign{\smallskip}
Cepheids & 18.52 \\
RR Lyraes & 18.49 \\
Miras & 18.48 \\
Eclipsers & 18.40: \\ 
Red Clump & 18.52 \\
SN1987A & 18.58 \\
 &  \\
Unweighted(6) & 18.50 \\
\noalign{\smallskip}
\tableline
\end{tabular}
}
\end{center}
\end{table}

Table 5 lists LMC distance moduli derived using various distance indicators.
The Cepheid modulus uses the mean zero-point of Table 1 together with a 
metallicity correction (see Feast 2003 for details). The Mira modulus uses the 
mean result of Table 4 together with the data of Feast et al. (1989).
The RR Lyrae modulus uses the zero-point of Table 3 together with the
LMC data of Clementini et al. (2003). The result for eclipsing
variables is from Ribas et al. (2002). The considerable spread in moduli of
the binaries studied is worrying (see Ribas et al. and Feast 2003).
The red clump modulus depends on absolute magnitudes derived from Hipparcos 
parallaxes of galactic clump stars. These require correction for both
age and metallicity effects and the quoted result depends on Girardi and
Salaris (2001) and Alves et al. (2002). The result from the ring round
SN1987A is from Panagia (1998).

A low value of the LMC Modulus ($18.35 \pm 0.05$) has been suggested from 
main sequence fitting to the young cluster NGC 1866 (Walker et al. 2002).
I have not included this since there are uncertainties affecting the
photometric calibration and reddening, and the need for a large
metallicity correction (see Feast 2003, 
Salaris et al. 2003, Groenewegen \& Salaris 2003). 

\section{Basic Distance Indicators and Cosmology}

Eight years ago when the Hipparcos parallaxes were released, they indicated
a zero-point for metal-normal Cepheids about 8 percent (0.17 mag) brighter
than that then being used by the HST Key project group
(Feast \& Catchpole 1997, Feast 1998). This difference
has now been virtually eliminated by two effects. The mean galactic
zero-point in Table 1 is 0.08 mag fainter than the 1997 parallax value,
whilst the key project team (Freedman et al. 2001) now apply a metallicity 
correct to their
LMC Cepheids which effectively increases their scale by 0.08 mag.

Whilst this basic agreement is heartening, it is important to bear
two things in mind.

(1) When one is trying to calibrate zero-points of Cepheids (or anything
else) at the 0.1mag level, all sorts of problems, some of which have been
mentioned above, arise.  These are currently being studied by a number of 
workers and one cannot realistically claim that they have all been solved.

(2) Whist we are getting agreement on the basic scale there remain 
significant differences in the values of $\rm H_{o}$ derived by the
Key programme group (Freedman et al. 2001)
who find $\rm 71\,km\,s^{-1}\,Mpc^{-1}$ (from SNIa) and the SNIa group 
(e.g. Tammmann et al. 2002) who find $\rm 60\,km\,s^{-1}\,Mpc^{-1}$. 
 using effectively the
same basic calibration. These difference arise to some extent in the
interpretation of HST data and may not be entirely solved till there
are more and better observations from space.

One may reasonably ask why one should want to improve on the Cepheid
based value of $\rm H_{o}$. Don`t we get a more accurate value from
a combination of WMAP with other surveys (e.g. 2dF)(Spergel et al. 2003, 
Lahav, this volume) and a $\rm \Lambda CDM$ model?
However the $\rm \Lambda CDM$ model is of such significance that it would
seem expedient to test it to the best of our ability, and one
way to do this is to derive an independent value of $\rm H_{o}$.

Another way to confront results from large scale structure is from 
estimates of the age of
the Universe, for which globular clusters give us a lower limit.
This too is partly a distance scale problem. 
A change in distance modulus of 0.1 mag results in a change of age of
about 1 Gyr. At present the best ages
are from main-sequence turn-offs with distance from main-sequence fitting
of subdwarfs. The evidence suggests that the most metal-poor
clusters are the oldest. This introduces a problem since there are
hardly any subdwarfs with suitable parallaxes that are as metal-poor
as the most metal-poor globular clusters.The best distance estimate 
for a metal-poor 
cluster is probably that of NGC 6397 by Gratton et al. (2003). They find 
the values 
listed in Table 6 without and with diffusion.
For an estimate of the age of the universe we have to add 
the epoch of cluster formation. If
cluster formation is dated to around z= 10 then we must add 0.5 Gyr.
Very recently two papers have discussed the effects on stellar models 
from a revision of the important
$\rm ^{14}N(p,\gamma)^{15}O$ rate (Imbriani et al. 2004, 
Degl`Innocenti et al. 2003). This work 
indicated that
globular cluster ages need increasing by between 0.7 and 1.0 Gyr.
If we adopt 0.8 Gyr for this correction, the final result for the age of
the Universe is 14.8 Gyr 
for a model including diffusion, compared with 13.7 $\pm$ 0.2 Gyr from a 
combination
of WMAP, 2dF etc. In view of the uncertainties in the globular
cluster age there is no evidence for disagreement but it 
whets ones appetite for better cluster ages. 
Whilst it is rather difficult to be sure of the uncertainties
in stellar evolutionary theory, much of the presently adopted estimates
of uncertainties in
globular cluster ages lies in the uncertainties in their derived
distances. Clearly work to improve these would be very valuable.

\begin{table}[!ht]
\caption{~~~~~~~~~~~~~~~The Globular Cluster NGC 6397}
\smallskip
\begin{center}
{\small
\begin{tabular}{ll}
\tableline
\noalign{\smallskip}
  & Age (Gyr) \\
\noalign{\smallskip}
\tableline
\noalign{\smallskip}
No Diffusion & $13.9 \pm 1.1$ \\
With Diffusion & $13.5 \pm 1.1$ \\
$\rm^{14}N(p,\gamma)^{15}O$ Correction & ~~0.8 \\
Corrected Age & 14.3 or 14.8 \\
z = 10 Correction & ~~0.5 \\
Estimated Age of Universe (NGC 6397)& 14.8 or 15.2 $\pm \geq 1.1$\\
Estimated Age of Universe (WMAP) & $13.4 \pm 0.3$ \\
Estimated Age of Universe (WMAP/2dF etc.) & $13.7 \pm 0.2$ \\
\noalign{\smallskip}
\tableline
\end{tabular}
}
\end{center}
\end{table}
\section{Acknowledgments}
I am grateful to Patricia Whitelock for information prior to publication.

\end{document}